\documentclass[aps,prb,superscriptaddress,amsfonts,amsmath,amssymb,showpacs,twocolumn,floatfix,nobalancelastpage]{revtex4-1}
\usepackage{url}
\usepackage{bm}
\usepackage{graphicx}
\usepackage{amsmath}
\usepackage{amstext}
\usepackage{amssymb}
\usepackage{amsfonts}
\usepackage{amsbsy}
\usepackage{verbatim}
\usepackage{color}
\usepackage[colorlinks=true, urlcolor=blue, linkcolor=blue, citecolor=blue, pdftex]{hyperref}
\usepackage{multirow}

\begin{document}

\title{Field-induced superfluids and Bose liquids in Projected Entangled Pair States}

\author{Didier \surname{Poilblanc} }
\affiliation{Laboratoire de Physique Th\'eorique, CNRS, UMR 5152 and Universit\'e de Toulouse, UPS,
F-31062 Toulouse, France}

\author{Norbert \surname{Schuch}}
\affiliation{Institut f\"ur Quanteninformation, RWTH Aachen, D-52056 Aachen, Germany}

\author{J.~Ignacio \surname{Cirac}}
\affiliation{Max-Planck-Institut f{\"{u}}r Quantenoptik,
Hans-Kopfermann-Str.\ 1, D-85748 Garching, Germany}



\date{\today}

\begin{abstract} 
In two-dimensional incompressible quantum spin liquids, a large enough magnetic field 
generically induces "doping" of polarized S=1 triplons or S=1/2 spinons.  We review a number of cases such as spin-3/2 AKLT or 
spin-1/2 Resonating Valence Bond (RVB) liquids where the Projected Entangled Pair States (PEPS) framework 
provides very simple and comprehensive pictures.
On the bipartite honeycomb lattice, simple PEPS can describe Bose condensed triplons (AKLT) or spinons (RVB) superfluids with transverse staggered (N\'eel) magnetic order.
On the Kagome lattice, doping the RVB state with deconfined spinons or triplons (i.e. spinon bound pairs)
yields uncondensed Bose liquids preserving $U(1)$ spin-rotation symmetry.
We find that spinon (triplon) doping destroys (preserves) the topological $\mathbb{Z}_2$ symmetry 
of the underlying RVB state. 
We also find that spinon doping induces longer range interactions in the entanglement Hamiltonian, suggesting the 
emergence of (additive) log-corrections to the entanglement entropy.

\end{abstract}
\maketitle

Magnetic frustration in quantum SU(2)-invariant spin systems of low
dimensionality -- typically in two dimensions (2D) - has the potential to
stabilize spin liquids with no magnetic order and gapped magnetic (i.e.
\hbox{spin-1} ``triplon") excitations. Two distinct important classes of
such states are Affleck-Kennedy-Lieb-Tasaki (AKLT)~\cite{AKLT} states and
{\it topological} spin liquids~\cite{wen91}.  The AKLT ground state (GS) is
simply constructed out of valence bonds (VB), is non-degenerate, and
breaks no symmetries (in 2D). The spin-3/2 AKLT state on the honeycomb
lattice has been proposed as a universal quantum computation
resource.~\cite{AKLT-spin32} Nearest neighbor (NN) {\it resonating}
valence bond (RVB) states~\cite{anderson} -- where neighboring spins 1/2
are paired up in resonating singlet dimers -- offer simple ans\"atze of a
new type of spin liquid, with $\mathbb{Z}_2$ symmetry on the kagome
lattice~\cite{rvb}.  A remarkable feature of gapped topological
($\mathbb{Z}_2$) liquid is that triplons spontaneously fractionalize into
deconfined spins 1/2 dubbed ``spinons". Other gapped (but spinless)
topological excitations are ``visons"~\cite{kivelson,senthil}, vortexlike
excitations which carry half a quantum of flux of the (underlying)
$\mathbb{Z}_2$ gauge field.  An external magnetic field plays the role of
a chemical potential both for the triplons or the spinons polarized along
the field, and hence controls their densities.  For liquids with fractional 
spinon excitations (which can only be created by pairs),
this issue has been investigating using simplified "doped" quantum dimer
models (QDM) representing a mixture of fluctuating dimers (mimicing
singlet VB)~\cite{rk} and mobile vacancies (representing
spinons)~\cite{qdm1}.  Despite their apparent simplicity, these models
exhibit very rich phase diagrams~\cite{qdm2} with i) superfluid (or
supersolid) phases -- breaking spontaneously the $U(1)$ symmetry
associated to the spin rotation around the magnetic field direction (equivalent to
spinon number conservation in these models)-- and ii) Bose or Fermi liquid
phases where the $U(1)$ symmetry is preserved. They also provide a
microscopic system where ``statistical transmutation"~\cite{qdm3} is
realized: vacancies can bind to visons and change their mutual statistics
(from bosons to fermions or vice versa) as originally proposed by
Kivelson~\cite{kivelson}. Such a scenario in a real quantum spin systems
has not been observed so far. 

\begin{figure}
\includegraphics[width=0.95\columnwidth]{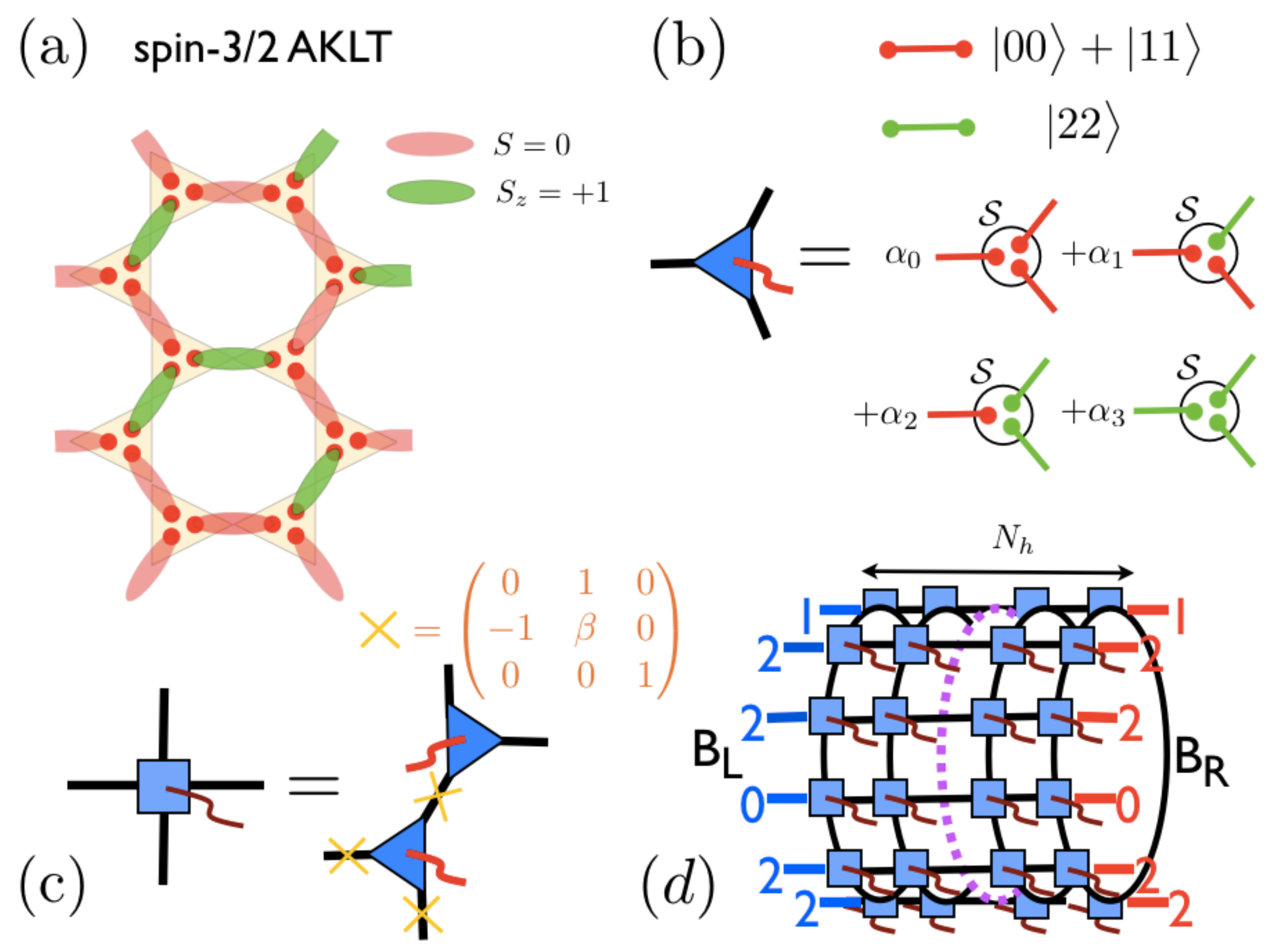}
\caption{(a) The honeycomb S=3/2 AKLT state under finite external magnetic
field.  Each spin is ``split" into three spins 1/2 (red dots). A ``valence
bond" configuration is constructed by pairing neighboring spins 1/2 into
singlets (red) or polarized-triplets (green).  (b) A $D=3$ generalization
of the AKLT PEPS is obtained by considering, in addition to the maximally
entangled states  $|00\rangle +|11\rangle$ on the bonds, new $|22\rangle$
bonds which represent triplets. Subsequently, the spins at each site are
symmetrized,  and a configuration with $p$ triplets is picked with weight
$\alpha_p$.  (c) The tensors on the A and B sublattices are grouped
together to form an effective square lattice. On the bonds, $X$ matrices
transform the $|00\rangle +|11\rangle$ states into $|01\rangle
-|10\rangle+\beta |11\rangle$.  (d) A cylinder geometry with boundaries
``vectors" $B_L$ and $B_R$ is used and the cylinder length $N_h$ is taken
to $\infty$. A vertical bipartition (dashed line) is used to compute the
boundary Hamiltonian and the entanglement spectrum.}
\label{Fig:peps}
\end{figure}

In recently synthesized Bi$_3$Mn$_4$O$_{12}$(NO$_3$) (named BiMnO), Heisenberg-like spin-3/2 
moments on the Mn$^{4+}$ ions form a bilayer honeycomb lattice~\cite{Mn_compound}. Despite the bipartite structure 
and the large antiferromagnetic {\it in-plane} coupling, BiMnO behaves as a spin liquid down to very low temperatures. 
A transition towards a N\'eel state (antiferromagnetic order) is induced by a moderate magnetic field~\cite{field_induced}. 
Theoretically, it has been suggested that the S=3/2 honeycomb AKLT model might well describe the properties of this material~\cite{spin32}
in zero and finite magnetic field. However the behavior under magnetic field of the S=3/2 AKLT model on the hexagonal lattice 
has not been investigated theoretically so far. 

In this work we investigate the behavior of  2D {\it gapped} spin liquids under an applied magnetic field. 
At small field before the
spin gap vanishes, the system remains in the same singlet GS. Therefore the magnetization curve 
(i.e. magnetization versus field) of spin-gapped systems 
generically starts with a zero-magnetization ``plateau" up to a lower critical field $h_{c,1}$ (which equals
the spin gap in appropriate units) before the magnetization starts to raise continuously. 
In other words, the zero-magnetization (gapped) phase has zero magnetic susceptibility and can therefore 
be viewed as an ``incompressible" liquid, in contrast to the compressible (i.e. gapless) finite magnetization phase
occurring at fields between $h_{c,1}$ and the saturation field $h_{c,2}$ above which the spin system is fully polarized. 
To describe the gapless phase, we construct Projected Entangled Pair States (PEPS)~\cite{peps,peps2,cirac} 
carrying a finite magnetization and originating from a simple PEPS representation of the corresponding zero-field 
spin liquid. Note that for fields below $h_{c,1}$ the PEPS representation of the GS is not changing.
The properties of these PEPS are investigated on infinitely long cylinders using standard methods~\cite{cirac}. 
By construction, these gapless spin liquid ans\"atze preserve full space group symmetry but may or may not break 
the U(1) spin-rotation symmetry around the magnetic field axis leading to (spin) superfluids or Bose liquids, respectively. 
Note that the translation symmetry-breaking ``crystals" leading to 
magnetization plateaux at special commensurate values~\cite{capponi,nishimoto} of the magnetization are not addressed here.

{\it AKLT state}:  we start with the S=3/2 AKLT Hamiltonian under an applied magnetic field:
\begin{equation}
H=H_{\rm AKLT}-hS^z=\sum_{\big< ij\big>} {\cal P}^{(S_T=3)}_{ij} - h\sum_{i=1}^NS_i^z
\end{equation}
where the sum is over all nearest neighbor (NN) bonds, ${\cal
P}^{(S_T=3)}_{ij}$ is the projector on total spin $S_T=3$ acting on the
product Hilbert spaces of sites $i$ and $j$, and $h=g\mu_B H$ is the
reduced effective field.  The AKLT ground state for $h=0$ can be
understood by viewing each spin-3/2 as being composed of three ``virtual''
spin-1/2 moments which are symmetrised on-site, with each spin-1/2 moment
forming a singlet with its neighbor; thus, it can be written exactly as a
simple \hbox{$D=2$} PEPS.  After turning on a finite $h$, the
magnetization $m=\big<\sum_i S^z_i\big>$ starts to rise above the
critical field $h_{c,1}$ for which the Zeeman energy overcomes the spin
energy gap.  Intuitively, an increasing density $x=m/m_{\rm sat}$ of singlets
is turned into polarized triplets until the saturation field $h_{c,2}$ is reached,
where all singlets are converted into triplets and $m=m_{\rm sat}=SN$,
where $N$ is the number of sites.  In such a picture, the triplets
"resonate" to gain energy and form a {\it resonating triplet bond} (RTB)
AKLT state. 
One way (i) to construct a RTB AKLT state is to extend the $D=2$ AKLT PEPS
to a $D=3$ PEPS with bonds $|01\rangle-|01\rangle+|22\rangle$,
where $|1\rangle$ and $|2\rangle$ are both assigned
$S_z=\tfrac12$ (i.e., $|22\rangle$ represents a triplet). The three
spins are then first symmetrized as in the AKLT state, and subsequently
projected onto a configuration with $p$ $|2\rangle$'s with relative weight
$\alpha_p$, as indicated in
Fig.~\ref{Fig:peps}(b).  We
expect the triplons to be weakly interacting, so that for simplicity, we
choose the coefficients $\alpha_p$ to be equal to their statistical
probabilities $\alpha_p^*(\lambda)$ with $\alpha_0^*=(1-\lambda)^3$,
$\alpha_1^*=3\lambda (1-\lambda^2)$, $\alpha_2^*=3\lambda^2 (1-\lambda)$
and $\alpha_3^*=\lambda^3$, depending on a single parameter $\lambda\in
[0,1]$ playing the role of a fugacity for the triplons. Another
(independent) way (ii) to introduce fluctuating $S_z=+1$ triplons is to
allow for an admixture of a triplet component on every bond, i.e., to replace
the virtual $|01\rangle -|10\rangle$ singlets by $|01\rangle -|10\rangle+\beta
|11\rangle$ states before the symmetrization, keeping the bond
dimension $D=2$.  Of course this admixture can be performed as well
on the extended $D=3$ state above, resulting in a two-parameter family of
PEPS.  In fact, both of these constructions can be understood as special
cases of a more general $9$-parameter RTB AKLT construction, as explained
in Appendix~\ref{app:triplon-peps}.  These states are generically not
invariant under $\exp{( i a S_z)}$ where $S_z$ is the total spin, and thus
can have a finite magnetization in the plane. 

\begin{figure}
\includegraphics[width=0.95\columnwidth]{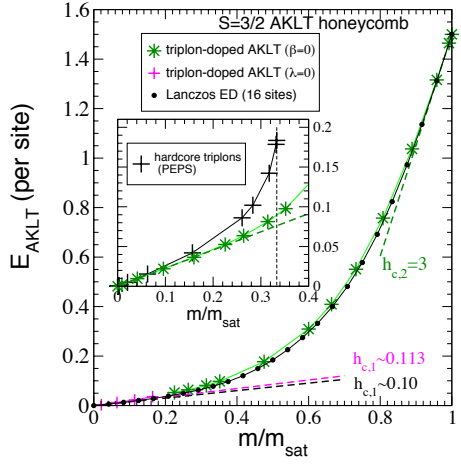}
\caption{Energy per site (different symbols are used when either $\lambda$ or $\beta$ is fixed to $0$) 
of the triplon-doped S=3/2 AKLT state
versus reduced magnetization computed on an infinite cylinder with perimeter $N_v=6$.  
The PEPS gives the {\it exact} asymptotic behavior close to saturation (i.e. slope $h_{c,2}=3$ at $m=m_{\rm sat}$) and the 
low-field slope ($h_{c,1}\simeq 0.113$) is in good agreement with Ref.~\onlinecite{spin32}.
Inset: comparison between the energies of AKLT PEPS with softcore and hardcore triplons. PEPS results are also compared 
with ED data~\protect\cite{sylvain}. }
\label{Fig:Energy}
\end{figure}

We have placed the square lattice of tensors on infinite cylinders with $N_v$ 
unit cells in the periodic (vertical) direction as shown in Fig.~\ref{Fig:peps}(d) and use standard techniques (involving 
exact tensor contractions and iterations of the transfer operator) to compute relevant observables.
We have investigated the variational energy $E_{\rm AKLT}(x)
=\frac{1}{N}\big< H_{\rm AKLT}\big>$ of the RTB-AKLT PEPS for $N_v=6$. Choosing
$\beta=0$ and varying $\lambda$ provides (approximately) the best energy
for $x>0.2$, while for $x<0.2$ the PEPS with $\lambda=0$ and $\beta\ne 0$ has lower energy.
The overall energy curve crudely obtained from these two separate PEPS is already quite 
accurate as shown in Fig.~\ref{Fig:Energy} when compared to Lanczos exact diagonalisations (ED)~\cite{sylvain}. 
By optimizing w.r.t. $\lambda$ and $\beta$ simultaneously, one can lower the energy even further down, especially
for $x< 0.3$ (see Appendix~\ref{app:spingap}).
The magnetization curve $m(h)/m_{\rm sat}$ can be obtained by minimizing 
$E_{\rm AKLT}(x) - (hS) x $ w.r.t. $x$.
The slopes $\frac{1}{S} dE_{\rm AKLT}/dx$ at $x=0$ and $x=1$ hence provide the lower and upper critical fields
$h_{c,1}$ and $h_{c,2}$ as indicated on Fig.~\ref{Fig:Energy}. Note that, in our units, $h_{c,1}$ equals the zero-field
spin gap $\Delta_S$. The physics close to saturation $m=m_{\rm sat}$ is captured 
exactly by our PEPS (with $\lambda\rightarrow 1$). Also, our estimate $h_{c,1}\simeq 0.113$ 
(for $\lambda,\beta\rightarrow 0$) is quite close to the extrapolated (zero-field)
spin gap $\Delta_S\simeq 0.10$~\cite{spin32}.

\begin{figure}
\includegraphics[width=0.95\columnwidth]{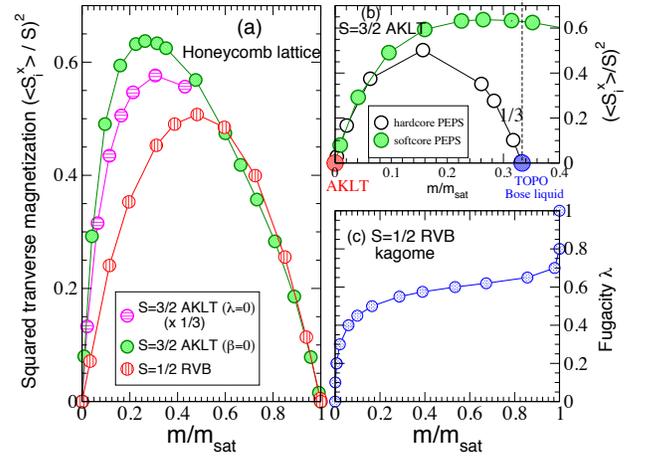}
\caption{Honeycomb lattice: square of the transverse magnetization $(\big< S_i^x\big>/S)^2$
versus reduced magnetization. All computations involve PEPS defined on an effective square lattice,
and are performed 
on an infinite cylinder with perimeter $N_v=6$. 
(a) Magnon-doped S=3/2 AKLT $D=2$ and $D=3$ states and $D=3$ spinon-doped S=1/2 NN RVB state. (b) Comparison between S=3/2 AKLT PEPS with doped softcore [as in (a)] and hardcore triplons. (c) Fugacity of the kagome RVB $D=3$ PEPS vs $m/m_{\rm sat}$.}
\label{Fig:Superfluid}
\end{figure}

We have found that, generically, the doped AKLT PEPS exhibit a transverse {\it staggered} magnetization in the plane
perpendicular to the field, i.e. $\big< S_i^x\big>=(-1)^i\times{\rm cst}$ (where $(-1)^i=\pm 1$ depending on the sublattice)
and $\big< S_i^y\big>=0$, hence breaking $U(1)$ symmetry.  This property is generic for any 
choice of the coefficients $\alpha_p$ and $\beta$ (except, possibly, for isolated points).
When $m\rightarrow 0$ ($m\rightarrow m_{\rm sat}$), the system can be understood in terms of a low concentration $x$
($1-x$) of interacting bosonic triplets (singlets) undergoing a Bose condensation and forming a correlated superfluid (SF). 
In the semi-classical approach applied to spins 1/2 forming dimers~\cite{spin_waves}, the quantity $(\big< S_i^x\big>/S)^2$
is the {\it condensate density} of triplons. We believe it is also a good (but approximate) indicator of Bose condensation for $S>1/2$, and plot it in Fig.~\ref{Fig:Superfluid}(a) versus 
the reduced magnetization for $\alpha_p=\alpha_p^*(\lambda)$, $\beta=0$ and for $\lambda=0$, $\beta\ne0$. 
When $x\rightarrow 1$, approaching saturation, $(\big< S_i^x\big>/S)^2\rightarrow 2(1-x)$ corresponding exactly to the effective singlet density.
In contrast, in the low magnetization limit $x\rightarrow 0$, $(\big< S_i^x\big>/S)^2\rightarrow 19x$ (for $\lambda=0$ and $\beta\ne0$ providing the best
ansatz).

We have also constructed a RTB-AKLT state with $\alpha_2=\alpha_3=0$, 
enforcing by hand an infinite repulsion between triplets. Interestingly, 
as seen in Fig.~\ref{Fig:Superfluid}(b), the SF order parameter $\big< S_i^x\big>$ now vanishes at exactly $m/m_{\rm sat}=1/3$
(again as a square root) giving rise  
to a (spin gapped) "Bose liquid" with restored U(1) symmetry. 
This state is the "negative" of the familiar S=1/2 (algebraic) RVB state: the singlet ($m=0$)  AKLT state
can be viewed as the new quantum "vacuum" where polarized hardcore triplets at 1/3-density resonate.
We believe that such a topological state (despite its poor variational energy for the AKLT Hamiltonian -- see Fig.~\ref{Fig:Energy}) 
could still be stabilized when the (effective) triplet repulsion is large enough (although not infinite). 

\begin{figure}
\includegraphics[width=0.95\columnwidth]{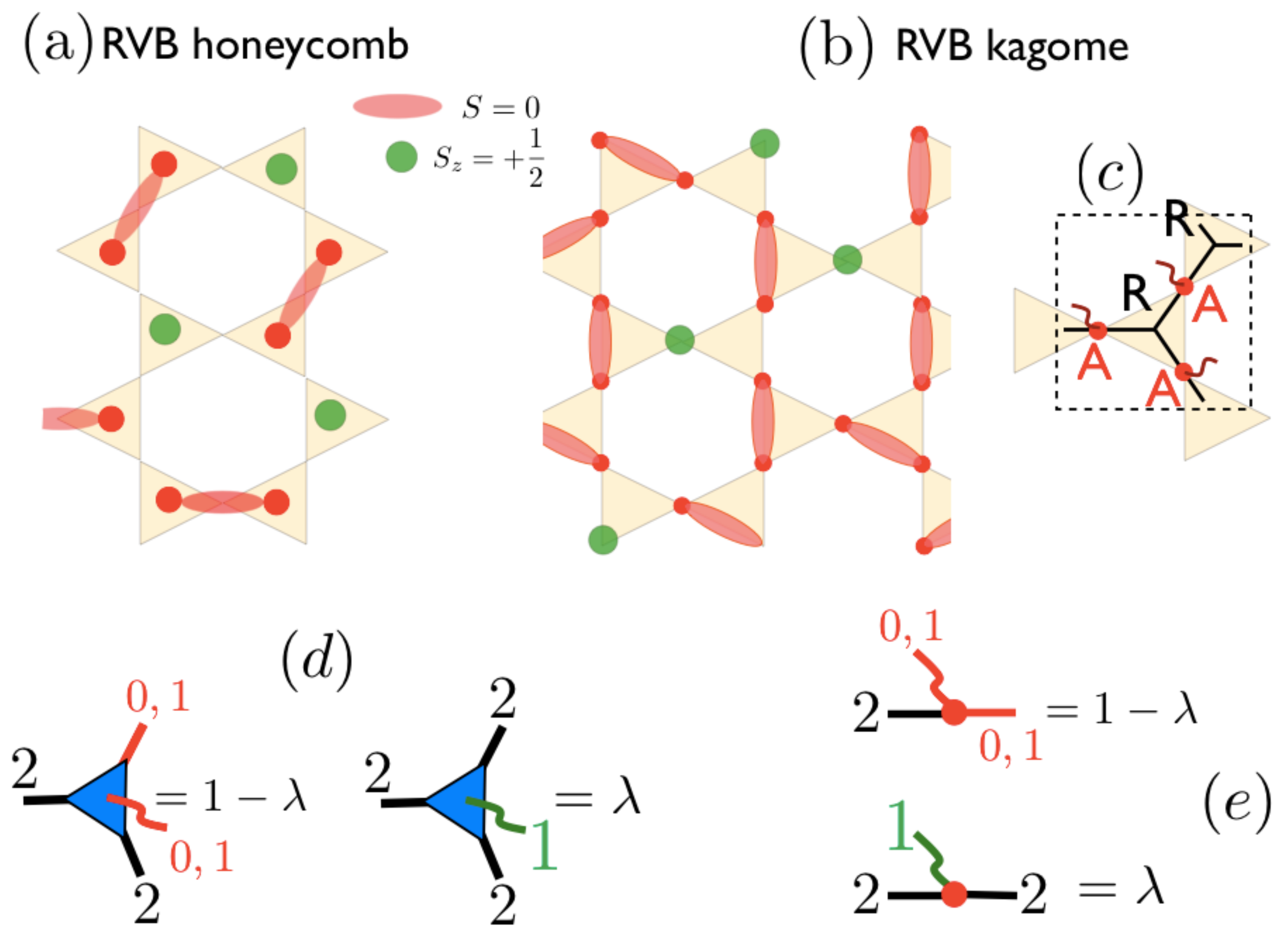}
\caption{Spinon-doped NN S=1/2 RVB wave functions on the honeycomb (a) and kagome (b) lattices.
An equal-weight superposition of all spinon (green dots) / VB (red ellipses) configurations is assumed. In the kagome RVB PEPS (c), 
three sites are grouped together (see Ref.~\protect\onlinecite{rvb} for details). Spinon doping in the honeycomb (d) and kagome (e) RVB PEPS is 
introduced by adding an extra non-zero tensor element. Its magnitude $\lambda$ plays the role of a fugacity for the spinons. }
\label{Fig:rvb}
\end{figure}

{\it RVB wave functions:}  The second class of wave functions we now investigate are NN S=1/2 RVB states, which we consider
on honeycomb and kagome lattices as depicted in
Fig.~\ref{Fig:rvb}(a,b). Both states have short range spin-spin correlations~\cite{anderson} but, on a bipartite lattice like the honeycomb lattice,
(singlet) dimer-dimer correlations are expected to be critical~\cite{rvb_topo,wang}. On the kagome lattice, the RVB state has $\mathbb{Z}_2$ 
topological structure~\cite{rvb,rvb_topo}. Interestingly, strong numerical evidence has been provided that the ground state of the NN S=1/2 quantum Heisenberg model (QHM) is indeed a gapped $\mathbb{Z}_2$ spin liquid~\cite{white,balents,schollwoeck}.

One of the most remarkable properties of such RVB states is that magnetic excitations are gapped deconfined S=1/2 spinons (marginally confined  on bipartite lattices) instead of S=1 triplons. Turning on a magnetic field larger that the spin gap, $h> \Delta_S$, will therefore dope spinons into the system. Simple extension of the RVB $D=3$ PEPS~\cite{rvb} can be realized to include a finite density of spinons as 
shown in Fig.~\ref{Fig:rvb}(c,d).  Note that incompressible phases such as those discussed in the literature~\cite{capponi,nishimoto}
for special fractional values of the magnetization (so-called "magnetization plateaux") are not addressed here. 

Computations are done on the same cylinder geometry (see Fig.~\ref{Fig:peps}(d)) as above. By increasing the fugacity $\lambda$ from $0$ to $1$, 
one can tune the magnetization between $0$ and $m_{\rm sat}$ 
as shown in Fig.~\ref{Fig:Superfluid}(c). As for the AKLT state, a finite transverse 
staggered magnetization is found for the honeycomb RVB state as shown in Fig.~\ref{Fig:Superfluid}(b).  Approaching saturation $m\rightarrow m_{\rm sat}$, the linear behaviors of the condensate density are identical, corresponding to the same condensation of singlets in a polarized ferromagnetic 
background. Above $h_{c,1}=\Delta_S$, the condensate density grows also linearly but the slope is smaller for the RVB state. 

A strikingly different behavior is found on the kagome lattice: no
transverse order is observed in the RVB PEPS, i.e.
$\big<S^i_x\big>=\big<S^i_y\big>=0$ (up to small finite size effects) for
all magnetizations. Therefore, the U(1) symmetry (spin rotation around
z-axis) is preserved so that this state can be viewed as a new type 
of gapless {\it Bose liquid}. 

We have also computed the variational energy of the QHM 
$H_{\rm Heis}=\sum_{\langle ij\rangle}\bf S_i\cdot S_j$. First,  the energy $E_{\rm QHM}(x)$ compares poorly to 
Lanczos ED~\cite{sylvain} and DMRG data~\cite{white} at low field and, in addition,  it has a 
slight negative curvature at $x\ll 1$ which signals an unphysical (small) jump in the magnetization curve
(see Appendix~\ref{app:spingap} for details). Our simple PEPS wave function then might not describe very well the physics of the QHM under magnetic field because 
i) the RVB state is only a poor ansatz for the ground state of the QHM at zero field~\cite{note1}. 
ii) Another source of difference might be that spinons could form bound NN pairs (triplons) in the QHM even though they are deconfined at long distance.  
We have tested this scenario by adding to the rank-3 $R$ tensors (see Fig.~\ref{Fig:rvb}(c)) the new non-zero elements $R(1,1,2)=R(1,2,1)=R(2,1,1)$
which control the density of $S_z=1$ triplons on NN sites. The triplon-doped RVB has 
indeed a lower energy than the spinon-doped RVB (see Appendix~\ref{app:spingap}) but the corresponding slope $\partial E/\partial x$ at $x=0$ 
remains too large compared to DMRG or ED.
Note that, when only triplons are doped, the $\mathbb{Z}_2$ topological sectors are preserved, since the doping keeps the $\mathbb{Z}_2$ symmetry of the tensors. 
iii) Thirdly, it is known from the studies of QDMs that spinons in $\mathbb{Z}_2$ spin liquids~\cite{qdm2,qdm3} 
can bind a topological vortex (vison) changing their mutual statistics from fermions to bosons, or vice versa. 
Further studies with more elaborate PEPS would be needed to investigate these possibilities. 

\begin{figure}
\includegraphics[width=0.95\columnwidth]{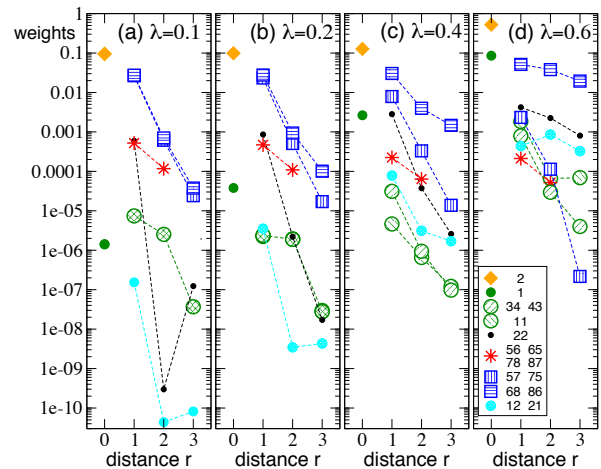}
\caption{Weights $|c_\nu|$ and $|d_{\nu\mu}(r)|^2$ of the one-body (i.e. $r=0$) and two-body operators
in the expansion of $H_b$ of the kagome spinon-doped RVB PEPS as a function of distance $r$, for increasing $\lambda$ values
(corresponding to reduced magnetization $x\sim10^{-3}, 0.007, 0.06, 0.53$, respectively).}
\label{Fig:bham}
\end{figure}

{\it Entanglement Hamiltonian~:} Entanglement measures offer new tools for characterizing exotic states like topological liquids.
If the (infinite) cylinder of Fig.~\ref{Fig:peps}(d) is partitioned in two $A$ and $B$ halves, the 2D reduced density matrix $\rho_A={\rm Tr}_B\{
|\Psi_{\rm PEPS}\big>\big<\Psi_{\rm PEPS}|\}$  of any $|\Psi_{\rm PEPS}\big>$ PEPS can be simply mapped, via a spectrum conserving 
isometry $U$, onto an operator $\sigma_b^2$ acting only on the $D^{\otimes N_v}$ edge (virtual) degrees of freedom~\cite{cirac}, i.e. $\rho_A=U^\dagger \sigma_b^2 \,U$.  
Therefore, it is convenient to define an {\it entanglement (or boundary) Hamiltonian} $H_b$ as $\sigma_b^2=\exp{(-H_b)}$. As $\sigma_b^2$, $H_b$
is one-dimensional and its spectrum -- the so-called {\it entanglement spectrum} (ES) -- is the same as the one of $-\ln{\rho_A}$.
In a $\mathbb{Z}_2$ topological liquid, the ES (and the associated $H_b$) depends on the choice of the boundary conditions $B_L(=B_R)$ 
-- due to the existence of two "even" and "odd" disconnected topological 
sectors -- and on the existence/absence of a $\mathbb{Z}_2$ flux through the cylinder~\cite{rvb_topo,topo_order}. 
Adding any magnetization in the PEPS
breaks topological order, since it break the gauge symmetry of the tensors
which is responsible for that. 
Therefore, at any (arbitrary small) doping, 
the two topological sectors are mixed and $H_b$ become independent of $B_L(=B_R)$ provided $N_h\rightarrow\infty$. 

In fact, the entanglement Hamiltonian of the $h=0$ RVB PEPS belongs to the $1/2\oplus 0$ representation
of SU(2) and its Hilbert space is the same as the one of a bosonic \hbox{t--J} model~\cite{rvb_topo}.
In the presence of a finite magnetization in the bulk, the SU(2) symmetry is broken but $H_b$ keeps 
the unbroken $U(1)$ symmetry of the bulk Bose liquid.
To have a better insight of the U(1) entanglement Hamiltonian, we expand it in terms of a basis of $M$-body operators, $M=0,1,2,\cdots$.
For this purpose, we use a local basis of $D^2=9$ (normalized)
${\hat x}_\nu$ operators, $\nu=0,\cdots,8$ which act 
on the local (i.e. at some site $i$) configurations $\{ |0\big>, |1\big>, |2\big> \}$, where 
$|2\big>$ is the vacuum or ``hole" state and $|0\big>$ and $|1\big>$ can be viewed as 
spin down and spin up states, respectively. More precisely,
${\hat x}_0=1^{\otimes 3}$, ${\hat x}_1=\sqrt{\frac{3}{2}}(|0\big>\big<0| -|1\big>\big<1|)$ 
and ${\hat x}_2=\frac{1}{\sqrt{2}}(|0\big>\big<0|+|1\big>\big<1|-2|2\big>\big<2|)$,
for the diagonal matrices, complemented by  
$\hat x_3=\hat x_4^\dagger=\sqrt{3}|0\big>\big<1|$
acting as (effective) spin-1/2 lowering/raising operators,
and $\hat x_5=\hat x_7^\dagger=\sqrt{3} |2\big>\big<0|$ and 
$\hat x_6=\hat x_8^\dagger=\sqrt{3} |2\big>\big<1|$ acting as hole hoppings. 
In this basis $H_b$ reads,
\begin{eqnarray}
H_b &=& c_0 N_v+\sum_{\nu,i} c_{\nu} {\hat x}_\nu^i 
+ \sum_{\nu,\mu,r,i} d_{\nu\mu}(r) \, {\hat x}_\nu^i {\hat x}_\mu^{i+r} 
+ \cdots \, \label{Eq:Hb0}
\end{eqnarray}
where site indices have been added and we restrict to the leading one-body and two-body terms.
The non-zero (real) coefficients in (\ref{Eq:Hb0}) computed 
on an infinitely-long cylinder of perimeter $N_v=6$ are shown in Fig.~\ref{Fig:bham}.
The leading hopping contributions are now split into two parts of different amplitudes, $d_{68}=d_{86}$ for the majority spins ($|1\big>$ states) larger than $d_{57}=d_{75}$ for the minority spins ($|0\big>$ states). The (very small) $h=0$ Heisenberg exchange ($d_{11}=d_{34}=d_{43}$) now takes the form of an anisotropic XXZ term ($d_{11}\ne d_{34}=d_{43}$). We also observe the emergence of new $U(1)$-invariant terms (forbidden by the $h=0$ SU(2) symmetry) e.g. an effective Zeeman term 
$\sum_i S_i^z$  of amplitude $d_1\ne 0$ and mixed terms like $\sum_{i} S_i^z (n_{i\pm r}-2/3)$ 
of amplitudes $d_{12}=d_{21}\ne 0$, where $n_i$ is the particle density (see corresponding ES in Appendix~\ref{app:ES}). 
For large enough magnetization, e.g. on Figs.~\ref{Fig:bham}(c,d), it becomes clear that the boundary Hamiltonian is long-range.
We believe that this is in fact a feature of all compressible (gapless) liquids for $x>0$, although the long-range "tails" 
are more difficult to detect numerically when $x\rightarrow 0$. 
Note also that, on bipartite lattices where transverse antiferromagnetic order is induced by the magnetic field, we found that $H_b$ 
does not conserve the total edge magnetization $S_z$ anymore.

To summarize, simple ($D=3$) PEPS have been constructed to understand
new phases induced by a magnetic field on various 2D magnetically 
disordered quantum spin systems. In the case of bi-partite lattices, 
our PEPS ans\"atze exhibit transverse (to the field) N\'eel order as e.g. in the 
S=3/2 AKLT and RVB states on the honeycomb lattice.
First, this confirms that the area law for the entanglement entropy can indeed occur  
in ground states with long-range magnetic order (here the entropy is bounded by $\ln{3}$ 
times the length of the cut).~\cite{EntrNeel}
Although, it would be possible to construct also U(1)-invariant PEPS on these lattices, such constructions
are much less intuitive. In addition, the variational energy of our polarized symmetry-broken 
PEPS is remarkably accurate in the case of the AKLT model. These facts strongly suggest 
that the existence of transverse staggered order is generic on such lattices (although exceptions could occur).
In the case of the topological kagome $\mathbb{Z}_2$ spin liquid (RVB state) 
doping by deconfined spinons or by triplons (spinon pairs) results in some new type of Bose 
liquids, i.e. gapless states preserving $U(1)$ $S_z$-symmetry. 
We note that such an uncondensed bosonic phase may have some interesting connections with the 
hoped for spin Bose-metal~\cite{spin_Bose_metal} with spinon "Bose surfaces"~\cite{note}.
The topological sectors of the $h=0$ gapped spin liquid are 
preserved when spinons form bound pairs (triplons) but
disappear for any small density of unbound spinons. 
We have computed the entanglement Hamiltonian of the spinon-doped RVB 
on the circumference of a bi-partitioned cylinder and showed that it becomes longer and longer 
range for increasing magnetization. 
Therefore we believe that {\it additive} logarithmic corrections to the entanglement 
entropy, as seen e.g. in numerical simulations of the N\'eel state~\cite{EntrNeel}
or in gapless spin liquids~\cite{wang}, 
are expected in all the compressible phases (i.e. for $0<x<1$) due to 
the long-range character of the corresponding entanglement Hamiltonians.
Lastly, investigating the energetics of our wave functions, we found that
the triplon-doped AKLT PEPS is a very good variational candidate for the AKLT
Hamiltonian with a Zeeman term. In contrast, our simple kagome RVB PEPS does
not seem to capture very well the effect of a magnetic field on the
spin-1/2 Heisenberg antiferromagnet; it may well be that spinons bind to
topological excitations of the $\mathbb{Z}_2$ liquid (visons) in order to
gain more kinetic energy. 
Another possibility is that some transverse magnetic order (e.g. non-collinear 3-sublattice order) may be more favorable energetically than
a U(1) symmetry-preserving Bose liquid. 
Let us add that being PEPS, all our trial
wavefunctions do appear as exact ground states of local parent
Hamiltonians,\cite{parentham} but in certain cases, these Hamiltonians are
not close to the parent Hamiltonians of the unperturbed
states.\cite{uncleham}

D.P.\ acknowledges partial supports by the ``Agence Nationale de
la Recherche" under grant No.~ANR~2010~BLANC~0406-0 and the
CALMIP supercomputer center (Toulouse).  N.S.\ acknowledges support by the
Alexander von Humboldt foundation.
D.P.\ is indebted to Sylvain Capponi for insightful  comments and for providing all
Lanczos exact diagonalization data shown in this paper.

Note added: after completion of this work we became aware of a related work 
on AKLT Hamiltonians by  Artur Garcia-Saez, Valentin Murg, and Tzu-Chieh Wei,
 arXiv:1308.3631.
 

\appendix

\section{\label{app:triplon-peps}
PEPS for resonating triplon AKLTs}

\newcommand{\ket}[1]{\vert#1\rangle}
\newcommand{\bra}[1]{\langle#1\vert}

In this appendix, we discuss the general form of a resonating triplon
doped AKLT, and explicitly explain how the constructions described in the
main text fit into this picture. To this end, we start from virtual
maximally entangled bonds of the form $\ket{01}-\ket{10}+\ket{22}$.   The
idea is that the $\{\ket{0},\ket{1}\}$ subspace holds the singlets, while
the $\ket{2}$ holds the triplets. Thus, both $\ket{1}$ and $\ket{2}$ are
understood to have $S_z=\tfrac12$, but they are distinguished by a
``triplet-ness'' quantum number $t$:
$\ket{1}\equiv\ket{S_z=\tfrac12,t=0}$, and $\ket{2}\equiv
\ket{S_z=\tfrac12,t=1}$.

To obtain a triplon-doped AKLT from these bonds, we now need to do two
things: (i) we need to symmetrize the virtual spins; (ii) we need to
choose the relative probabilities for having a certain number $p$ of
triplets at each site and subsequently erase the ``triplet-ness'' quantum
number, i.e., make $\ket{1}$ and $\ket{2}$ indistinguishable.  These steps
can be carried out in either ordering, which will generally give different
outcomes (since converting $\ket{2}$ to $\ket{1}$ changes the norm of
vectors).  In the most general framework, this can be expressed by
decomposing the total PEPS projector $\mathcal P$ as (i) a symmetrization
map 
\begin{equation}
    \label{eq:appc:sym-map}
\mathcal S =\!\!\! 
\sum_{m=-\frac32}^{\frac32}\sum_{p=0}^{\frac{3}{2}+m}
    w_{m,p}\ket{S_z=m;t=p} \left[\sum \bra{i_1,i_2,i_3} \right]
\end{equation}
where the r.h.s.\ sum symmetrizes over all $\bra{i_1,i_2,i_3}$  with
$\tfrac32-m$ $0$'s (i.e., $S_z=m$) and $p$ $2$'s; and (ii) a map
projecting onto a given relative weight of different triplet numbers
$t=p$, 
\begin{equation}
    \label{eq:appc:weightproj-map}
\mathcal T = 
\sum_{p=0}^3 
\gamma_p
\sum_{m=p-\frac32}^{\frac32} 
\ket{S_z=m}\bra{S_z=m;t=p}\ ,
\end{equation}
such that $\mathcal P=\mathcal T\,\mathcal S$. Clearly, the $\gamma_p$ can
be absorbed into the $w_{m,p}$, leaving us with a $(10-1)=9$-parameter family of
PEPS.

We will now show that the families of triplon-doped AKLT PEPS studied in
the paper both fall into this family.  The first one (with $\beta=0$) is
obtained by choosing the symmetrization map $\mathcal S$ to be a
projector, i.e., the $w_{m,p}$ are equal to the square root of the number
of terms in the r.h.s.\
sum in (\ref{eq:appc:sym-map}),
\begin{align*}
&w_{3/2,0}=w_{3/2,3}=1\,, &&  
    w_{3/2,1}=w_{3/2,2}=\tfrac1{\sqrt{3}}\,,\\
&w_{1/2,0}=w_{1/2,2}=\tfrac1{\sqrt{3}}\,, &&  
    w_{1/2,1}=\tfrac1{\sqrt{6}}\,,\\
&w_{-1/2,0}=w_{-1/2,1}=\tfrac{1}{\sqrt{3}}\,, &&
    w_{-3/2,0}=1\,,
\end{align*}
and by setting $\gamma_p=\alpha_p$ in (\ref{eq:appc:weightproj-map}), 
leaving us with the four parameter family described in the main text.

The second variant described in the main text -- using a bond
$\ket{01}-\ket{10}+\beta \ket{11}$, while leaving the $\alpha_p=0$ for
$p>0$ -- can
be understood as first using the $\ket{2}$ level to pick a triplet on each
bond with weight $\sqrt{\beta}$ on each end, turning the $\ket{2}$ into a
$\ket{1}$, and subsequently projecting on the symmetric subspace.  Since
turning the $\ket{2}$ into a $\ket{1}$ effectively changes the number of
terms in the r.h.s.\ sum of (\ref{eq:appc:sym-map}), this leads to
different weights $w_{m,t}$, and one finds that this PEPS can be described
by choosing $w_{\pm 3/2,p}=1$, $w_{\pm 1/2,p}=1/\sqrt{3}$ in
(\ref{eq:appc:sym-map}), and $\gamma_p=\beta^{p/2}$ in
(\ref{eq:appc:weightproj-map}).

Clearly, due to linearity the ansatz where both the $\alpha_p$ and $\beta$
are non-zero can also be described using the full family defined by
Eqs.~(\ref{eq:appc:sym-map}) and (\ref{eq:appc:weightproj-map}).

\section{\label{app:spingap}
Spin gaps and magnetization jumps}

\begin{figure}
\bigskip
\includegraphics[width=0.95\columnwidth]{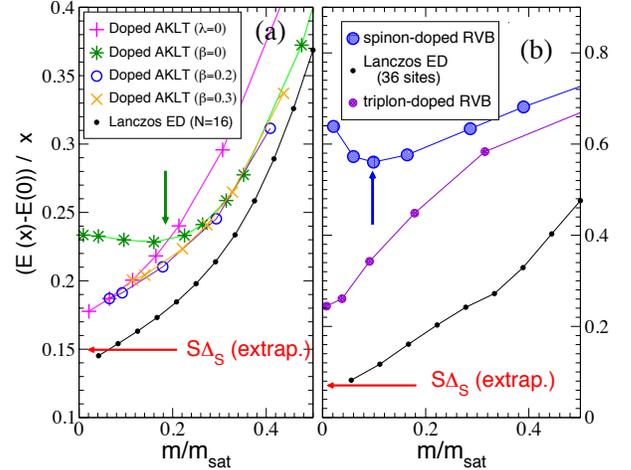}
\caption{$(E(x)-E(0))/x$ versus $x=m/m_{\rm sat}$ for small $x$ compared to ED data~\cite{sylvain}
for several honeycomb doped AKLT (a) and kagome RVB (b) states on $N_v=6$ cylinders. Shallow minima (if any)
are shown by vertical arrows. The estimated extrapolated spin gaps of Refs.~\onlinecite{spin32} 
and \onlinecite{white} (multiplied by the spin $S$) -- to be compared with the $m/m_{\rm sat}\rightarrow 0$ limits of the various curves -- 
are shown by horizontal (red) arrows.  }
\label{Fig:ps}
\end{figure}

The derivative $\partial E/\partial x|_{x=0}$ provides an estimate of $S\Delta_S$, the spin gap (times $S$) of the incompressible phase at $x=0$.
We have plotted  the quantity $(E(x)-E(0)/x)$ in Fig.~\ref{Fig:ps} for doped $S=3/2$ AKLT and RVB states which gives the above derivative when 
taking the limit $x\rightarrow 0$.
The agreement with the gap estimate of the S=3/2 AKLT model~\cite{spin32} is fairly good for the triplon-doped AKLT state,
with a proper choice of the parameters. 
In contrast, the spinon-doped RVB state and, to a lesser extent, the triplon-doped RVB state overestimates the small 
gap of the kagome quantum antiferromagnet by a large amount. 

A close look at the variational PEPS energies $E(x)$ reveals a small negative curvature 
when \hbox{$x\rightarrow 0$}, both for the $\beta=0$ AKLT and the
spinon-doped kagome RVB states. 
$\partial^2 E / \partial x^2$ is the inverse spin susceptibility $\chi^{-1}$ and $\chi^{-1}<0$ signals a (weak) instability towards phase separation
between a $x=0$ phase and a phase with $x=x_{\rm min}$ where $x_{\rm min}$ is given by the minimum of $(E(x)-E(0)/x)$ (Maxwell construction). 
When increasing the field $h$, $x(h)$ then jumps from $0$ to
$x_{\rm min}$ at $h=h_{\rm c,1}$.
As shown in Fig.~\ref{Fig:ps} $x_{\rm min}\sim 0.2$ ($x_{\rm min}\sim 0.1$) for the AKLT state (RVB state) on the (infinite) $N_v=6$ cylinder. 
However, comparison with Lanczos ED suggests that this is in fact a finite size effect of the 
variational ans\"atze on finite cylinder. More precisely, comparing $N_v=4$ and $N_v=6$ cylinders
we get $x_{\rm min}\sim 1/N_v$ suggesting that phase separation disappears when $N_v\rightarrow\infty$. 
Note that, in contrast, the $\beta\ne 0$ AKLT state and kagome triplon-doped RVB (with improved variational energies) 
do not show phase separation for finite $N_v$. 

\section{\label{app:ES}
Entanglement spectrum}

\begin{figure}
\includegraphics[width=0.95\columnwidth]{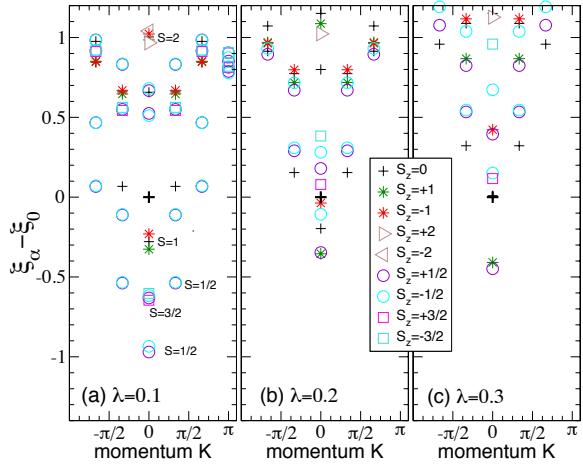}
\caption{Entanglement spectrum of a bi-partitioned $N_v=6$ kagome RVB cylinder as a function of the 
momentum along the cut, for different values of the spinon fugacity $\lambda$ corresponding to
$m/m_{\rm sat}\sim10^{-3}, 0.007, 0.02$, respectively. Different symbols are used for different $S_z$ sectors of the edge.
The same $S_z=0$, $K=0$ state is used as energy reference $\xi_0$ (bold $+$ symbol).  }
\label{Fig:es}
\end{figure}

By definition the ES is the spectrum of $-\ln{\rho_A}$. Since $\rho_A$ and $\sigma_b^2=\exp{(-H_b)}$ are related by an isometry, it is also the spectrum of the boundary Hamiltonian $H_b$. ES are shown in Fig.~\ref{Fig:es} 
for 3 values of the fugacity $\lambda$, as a function of the momentum along the cut. 
The $h=0$ SU(2) spin multiplets are split by an arbitrary small spinon concentration as shown 
in Fig.~\ref{Fig:es}(a). For increasing $\lambda$ (and magnetization), the splittings of the Kramers multiplets 
increase (see Fig.~\ref{Fig:es}(b,c)) due to the relative increase of the amplitudes of an effective 
Zeeman term and of new SU(2)-symmetry breaking many-body terms in the boundary Hamiltonian (see main text). 

\end{document}